\documentstyle[12lomcon,cite]{article}

\begin{document}

\title{COLLAPSE OF POSITRONIUM AND VACUUM INSTABILITY}

\author{A.E. Shabad \footnote{e-mail: shabad@lpi.ru}},

\address{P.N. Lebedev Physics Institute, Leninsky prospekt 53, Moscow 119991, Russia}

\author{V.V. Usov\footnote{e-mail: fnusov@wicc.weizman.ac.il}}

\address{Center for Astrophysics, Weizmann Institute
of Science, Rehovot 76100, Israel}

\maketitle

\abstracts{A hypercritical value for the magnetic field is
determined, which provides the full compensation of the
positronium rest mass by the binding energy in the maximum
symmetry state and disappearance of the energy gap separating the
electron-positron system from the vacuum. The compensation becomes
possible owing to the falling to the center phenomenon. The
structure of the vacuum is described in terms of strongly
localized states of tightly mutually bound (or confined) pairs.
Their delocalization for still higher magnetic field, capable of
screening its further growth, is discussed.}

It is accepted that magnetic fields are stable in pure quantum
electrodynamics (QED), and other interactions (weak or strong) or
magnetic monopoles have to be involved to make the magnetized
vacuum unstable \cite{NO78}. In this Talk (see Ref. \cite{SU05}
for a detailed version) we establish, within the frame of QED,
that there exists a hypercritical value of the magnetic field
\begin{eqnarray}\label{final}
B^{(1)}_{\rm hpcr}\simeq \frac{m^2}{4e}\exp\left\{\frac{\pi^{3/2}}
{\sqrt{\alpha}}+2C_{\rm E}\right\}\simeq 10^{28}B_0\simeq
10^{42}\rm G\,.
\end{eqnarray}

Here $\alpha=e^2/4\pi\simeq 1/137,$ $B_0=m^2/e=4.4\times
10^{13}$G, $m$ is the electron mass. The value (\ref{final}) is
less than the magnetic field $\sim (10^{47}-10^{48})$ G expected
to be present near superconductive cosmic strings \cite{W95} and
the one ($\sim 10^{47}$G) produced at the beginning of the
inflation \cite{KK89}.
 The hypercritical magnetic field leads to the shrinking of the
energy gap between an electron and positron that takes place due
to the falling to the center. The latter is caused by the
ultraviolet singularity of the photon propagator (or Coulomb
potential) \cite{goldstein} plus the dimensional reduction in the
magnetic field \cite{loudon} - \cite{gusynin}. We discuss the
vacuum structure around the hypercritical magnetic field and its
possible decay under a further growth of the magnetic field - that
may cause its screening.

We rely on  the theory of the falling to the center developed in
\cite{shabad} that implies deviations from the standard quantum
theory manifesting themselves when extremely large electric fields
near the  singularity become important. In that theory the
singularity in the Schr\"{o}dinger-like equation yields a singular
measure in the scalar product and hence the geometry of a
black-hole-like object. (Stress, that the geometry is induced: no
interaction of  gravitational origin is included.)

We proceed from the (3+1)-dimensional Bethe-Salpeter (BS) equation
in an approximation, which is the ladder approximation once the
photon propagator (in the coordinate space) is taken in the
Feynman gauge: $D_{ij}(x)=g_{ij}D(x^2)$, $x^2\equiv x_0^2-{\bf
x}^2$, $g_{ii}=(1,-1-1-1)$. In an asymptotically strong magnetic
field this equation may be written \cite{SU05} in the following
(1+1)-form  covariant under the Lorentz transformations along the
axis 3:
\begin{eqnarray}\label{closed1}({{\rm i}\overrightarrow{\hat{
\partial_\|}}- \frac{\hat{P_\parallel}}2-m})
\Theta(t,z)(-{\rm i}\overleftarrow{\hat{
\partial_\parallel}}- \frac{\hat{P_\parallel}}2-m)\quad\quad\quad
\nonumber\\ = {\rm i}
8\pi\alpha~\sum_{i=0,3}D\left(t^2-z^2-\frac{P_\perp^2}{(eB)^2}\right)g_{ii}\gamma_i
\Theta(t,z)\gamma_i,
\end{eqnarray}
where $\Theta (t,z)$ is the 4$\times 4$ (Ritus transform of) BS
amplitude, $t=x^{\rm e}_0- x^{\rm p}_0$ and $z=x^{\rm e}_3-x^{\rm
p}_3$ are the differences of the coordinates of the electron (e)
and positron (p) along the time $x_0$ and along the magnetic field
${\bf B}=(0,0,B_3=B)$. $P_\|$ and $P_\perp$ are projections of the
total (generalized) momentum of the positronium onto the (0,3)-
subspace and the (1,2)-subspace. Only two Dirac gamma-matrices,
$\gamma_{0,3},$ are involved, $\hat{
\partial_\|}=\partial_t\gamma_0-
\partial_z\gamma_3$, $\hat{P_\|}=P_0\gamma_0-P_3\gamma_3$.

Equation (\ref{closed1}) is valid in the  domain, where the
argument of  $D$ is greater than the electron  Larmour radius
squared $(L_B)^2=(eB)^{-1}$. When $B=\infty$, this domain covers
the whole exterior of the light cone $z^2-t^2\geq 0$.The argument
of the original photon propagator $(x^{\rm e}-x^{\rm p})^2$ has
proved to be replaced in (\ref{closed1}) by
$t^2-z^2-(\widetilde{x}_\perp^{\rm e}-\widetilde{x}_\perp^{\rm
p})^2 =t^2-z^2-P_\perp^2/(eB)^2$, where $\widetilde{x}_\perp^{\rm
e,p}$ are the center of orbit coordinates of the two particles in
the transversal plane. Now that after the dimensional reduction
this subspace no longer exists these substitute for the
transversal particle coordinates themselves:
$\widetilde{x}_\perp^{\rm e,p}$are not coordinates, but quantum
numbers of the transverse momenta. The mechanism of replacement of
a coordinate by a quantum number is the same as 
in \cite{ShUs}.

In deriving equation (\ref{closed1}) the expansion over the
complete set of Ritus matrix eigenfunctions \cite{ritus} was used
in \cite{SU05} that accumulate the dependence on the transversal
spacial and spinorial degrees of freedom. This expansion yields an
infinite set of equations, where different pairs of Landau quantum
numbers $n^{\rm e}$, $n^{\rm p}$ are entangled, Eq.
(\ref{closed1}) being the equation for the ($n^{\rm e}=n^{\rm
p}=0$)-component that decouples from this set in the limit
$B=\infty$.

In the ultra-relativistic limit $P_0=P_3=P_\perp=0$ equation
(\ref{closed1}) is solved by the most symmetric Ansatz
$\Theta=I\Phi$, where $I$ is the unit matrix.

The ultraviolet singularity  on the light cone ($x^2=0$) of the
free photon propagator, $D(x^2)=-(\rm i/4\pi^2)(1$/$x^2)$, after
this expression is used in
  eq. (\ref{closed1}),  leads to falling to the center in the
Schr$\ddot{\rm o}$dinger-like
 differential equation\begin{eqnarray}\label{last3} -\frac{{\rm
d}^2\Psi(s)}{{\rm d} s^2}+\left(m^2-\frac
1{4s^2}\right)\Psi(s)=\frac{4\alpha}{\pi}\frac 1{s^2}\Psi(s),\quad
(eB)^{-1/2}\ll s\leq \infty,
\end{eqnarray}
to which the radial part of equation (\ref{closed1}) is reduced in
the most
 symmetrical case, when the wave function $\Phi
(x)=s^{-1/2}\Psi(s)$ does not depend on the hyperbolic angle
$\phi$ in the space-like region of the two-dimensional Minkowsky
space, $t=s\sinh\phi,\; z=s\cosh\phi,\;s=\sqrt{z^2-t^2}$.

The solution that decreases at $s\rightarrow \infty$ is given by
the McDonald function with imaginary index:
$\Psi(s)=\sqrt{s}\;K_\nu(ms),\quad \nu= {\rm
i}\,2\sqrt{\alpha/\pi}\simeq 0.096\,{\rm i}\,,$ oscillating when
$s\rightarrow 0.$ The falling to the center \cite{QM} holds for
any positive $\alpha$.

According to \cite{shabad}, the singular equation (\ref{last3})
should be considered as the generalized eigenvalue problem with
respect to $\alpha$. The operator in the left-hand side is
self-adjoint provided the standing wave condition is imposed,
\begin{eqnarray}\label{stand} \left.\Psi (s)\right |_{s=L_B}=0 \,,
\end{eqnarray} that treats the Larmour radius as the lower
edge of the normalization box. The discrete eigenvalues
$\alpha_n(L_B)$ condense in the limit $B=\infty$ to become a
continuum of states that form the (rigged) Hilbert space of
vectors orthogonal with the singular measure $s^{-2}{\rm d}s$. The
latter fact allows to normalize them to $\delta$-functions and
 interpret as free particles emitted and
absorbed by the singular center. As long as the Larmour radius
is much smaller than the only characteristic length in
eq.(\ref{last3}), electron Compton length, $L_B\ll m^{-1}\simeq
3.9\times 10^{-11}$ cm., the small-distance asymptotic regime is
reached, and nothing "behind the horizon," $s<L_B,$ - where the
two-dimensional equations (\ref{closed1})
 and (\ref{last3}) are not valid - may affect the
problem. In this way the existence of the limit of vanishing
regularization length, impossible in the standard theory, is
achieved.

For sufficiently large magnetic fields $L_B$ becomes so small that
the region where $K_\nu(ms)$ oscillates gets inside the domain of
validity of equation (\ref{last3}). The value of the magnetic
field when this  happens for the first node is just (\ref{final}).
The corresponding Larmour radius is about fourteen orders of
magnitude smaller than $m^{-1}$ and makes $\sim 10^{-25}$ cm. Eq.
(\ref{final}) tells how large the magnetic field should be in
order that the boundary problem (\ref{last3}), (\ref{stand}) might
have a solution, in other words, that the point $P_0={\bf P}=0$
might belong to the spectrum. Therefore, if the magnetic field
exceeds the first hypercritical value  the positronium ground
state exists \footnote{A relation like (\ref{final}) is present in
\cite{gusynin}. There, however, a different problem is studied
and, correspondingly, a different meaning is attributed to that
relation: it expresses the mass gained dynamically - in the course
of spontaneous breakdown of the chiral invariance in massless QED
- by a massless Fermion in terms of the magnetic field applied to
it. Later, in \cite{gusynin2} the authors revised that relation in
favor of a different approximation. Supposedly, the revised
relation may be of use in the problem of ultimate magnetic field
dealt with here.} with its rest energy compensated for by the mass
defect.

The ultra-relativistic state $P_\mu=0$ has the internal structure
of what was called a "confined state", belonging to kinematical
domain called "sector III" in \cite{shabad}, i.e. the one whose
wave function behaves as a standing wave combination of free
particles near the lower edge of the normalization box and
decreases as $\exp (-ms)$ at large distances. The effective "Bohr
radius", i.e. the value of $s$ that provides the maximum to the
wave function  makes $s_{\rm max}=0.17 m^{-1}$. This is certainly
much less than the standard Bohr radius $(e^2m)^{-1}$. Taken at
the level of 1/2 of its maximum value, the wave function is
concentrated within the limits  $0.006 ~m^{-1}<s<1.1~ m^{-1}$. But
the effective region occupied by the confined state is still much
closer to $s=0$, since - in accord with the aforesaid - the
probability density of the confined state is the wave function
squared $weighted ~with~ the~ measure$ $s^{-2}{\rm d} s$
$singular~ in~ the~ origin$ \cite{shabad} and is hence
concentrated near the edge of the normalization box $s\simeq
10^{-25}$cm, and not in the vicinity of the maximum of the wave
function. The electric fields at such distances are about
$10^{43}$ Volt/cm. Certainly, there is no evidence that the
standard quantum theory should be valid under such conditions.
This fact encourages the use of a theory that admits deviations
from the standard quantum theory that close to the singularity
point.

At $B=B^{(1)}_{\rm hpcr}$ the total energy and momentum of a
positronium in the ground state are zero. This state is not
separated from the vacuum by an energy gap, and has maximum
symmetry in the coordinate and spin space. Hence, it may be
related to the vacuum and responsible for its structure.

 At $B>B^{(1)}_{\rm hpcr}$ the
eigenvalues of the BS equation (\ref{closed1}) for the total
2-momentum components $P_{0,3}$ of the e$^+$-e$^-$ system are
expected to shift into the space-like region (we keep
$P_\perp=0$), whereas for $B<B_{\rm hpcr}^{(1)}$ the c.m.
2-momentum of the real pair was, naturally, time-like. If
$P_{0,3}\neq 0,$ at least for far space-like $P_\parallel$, the
situation can be modelled by the same equation as (\ref{last3}),
but with the large negative quantity $~~m^2+P_\|^2/4~~$
substituted for  $m^2$. Then the wave function would contain two
oscillating exponentials for large space-like intervals,
\begin{eqnarray}\label{hankel} \exp\left\{\pm{\rm i}
s\sqrt{\left|m^2+\frac{P_\|^2}4\right|}\right\}\exp\left\{{\rm i}
P_\|\frac{x_{\rm e}+x_{\rm p}}2\right\},\quad  s\gg
(-P_\|^2)^{-1/2},\end{eqnarray} and two oscillating exponentials
$\exp (\pm2{\rm i}\ln s\,\sqrt{\alpha/\pi})$ for small ones,
$s\sim L_{\rm B}$. In the Lorentz frame, where $P_0=0, P_3\neq 0$,
and with the time arguments in the two-time BS amplitude equal to
one another: $x_0^{\rm e}=x_0^{\rm p}$, the solution oscillates
along the magnetic field with respect to the relative coordinate
$x_3^{\rm e}-x_3^{\rm p}$ (mutually free particles) and with
respect to the c.m. coordinate $x_3^{\rm e}+x_3^{\rm p}$ (vacuum
lattice).

We are now in the kinematical domain called sector IV, or
deconfinement sector in Refs.\cite{shabad}. Here the constituents
are free at large intervals and near the singular point $s=0$. The
wave incoming from infinity is partially reflected, and partially
penetrates to the singular point, the probability of creation of
the delocalized (free) states being determined by the barrier
transmission coefficient \cite{shabad}. Such states may exist if
one succeeds to satisfy e.g. periodic conditions, to be imposed on
the lower and upper boundaries of the normalization volume,
instead of  condition (\ref{stand}), appropriate in sector III.
The possibility to obey them is provided again by the falling to
the center. Now, the delocalized states in two-dimensional
Minkowsky space correspond to electron and positron that circle
along Larmour orbits with vanishing radii in the plane orthogonal
to the magnetic field and simultaneously perform, when the
interval between them is large, a free motion along the magnetic
field. They have magnetic moments and seem to be capable of
screening the magnetic field. This provides the mechanism that may
prevent the classical magnetic field from being larger than  a
{\it second hypercritical field}, for which the delocalization
first appears. No sooner than the delocalized states are found in
our present problem one may definitely claim the instability of
the vacuum with the second hypercritical magnetic field or - which
is the same - the instability of such field under the pair
creation that might provide the mechanism for its diminishing. For
the present, we state that the first hypervalue (\ref{final}) is
such a value of the magnetic field, the exceeding of which would
already cause restructuring of the vacuum.

\section*{Acknowledgments}
This work was supported by the Russian
 Foundation for Basic Research
 (project no 05-02-17217) and the President of Russia Programme
   (LSS-1578.2003.2), as well as
by the Israel Science Foundation of the Israel Academy of Sciences
and Humanities.

\end{document}